\documentclass[sigconf,natbib=true,balance=true]{acmart}

\usepackage{listings}
\usepackage{adjustbox}

\usepackage[show]{chato-notes}
\usepackage{listings}
\usepackage{xcolor}
\usepackage{multirow}
\usepackage{float}
\usepackage{flushend}
\usepackage{balance}

\usepackage{listings}
\AtBeginDocument{%
  \providecommand\BibTeX{{%
    \normalfont B\kern-0.5em{\scshape i\kern-0.25em b}\kern-0.8em\TeX}}}

\setcopyright{acmcopyright}
\copyrightyear{2024}
\acmYear{2024}
\acmDOI{XXXXXXX.XXXXXXX}

\acmConference[IR-RAG @ SIGIR '24]{The 47\textsuperscript{th} International ACM SIGIR Conference on Research and Development in Information Retrieval}{July 14--18, 2024}{Washington D.C., USA}
\acmBooktitle{IR-RAG @ SIGIR '24: The 47\textsuperscript{th} International ACM SIGIR Conference on Research and Development in Information Retrieval,
 July 14--18, 2024, Washington D.C., USA} 
\acmPrice{15.00}
\acmISBN{978-1-4503-XXXX-X/18/06}

\begin{document}

\title{Generating Query Recommendations via LLMs}

\author{Andrea Bacciu}
\orcid{0009-0007-1322-343X}
\affiliation{%
\institution{Sapienza University\includegraphics[height=2ex]{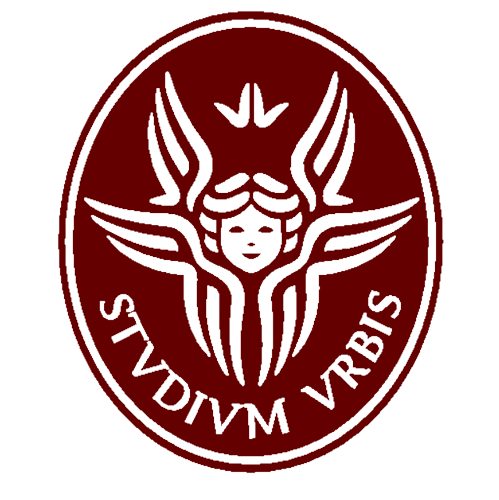}}
        \city{Rome}
        \country{Italy}
}
\email{bacciu@diag.uniroma1.it}
\authornote{Work done before joining Amazon.}

\author{Enrico Palumbo}
\orcid{0000-0003-3898-7480}
\affiliation{%
\institution{Spotify \includegraphics[height=2ex]{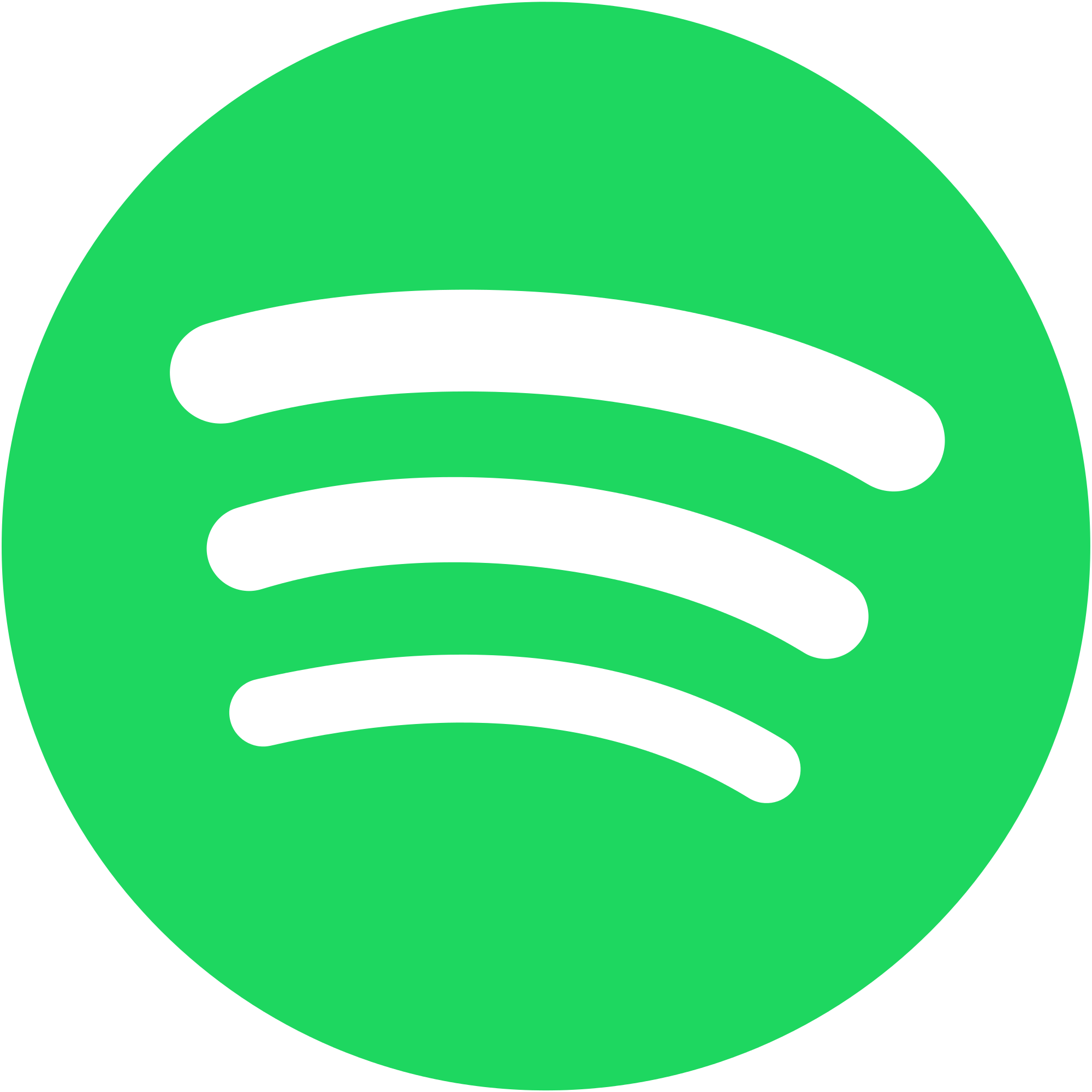}}
        \city{Turin}
        \country{Italy}
}

\author{Andreas Damianou}
\orcid{0000-0001-9383-5100}
\affiliation{%
\institution{Spotify \includegraphics[height=2ex]{images/logo.png}}
\city{Cambridge}
\country{United Kingdom}
}
\email{andreasd@spotify.com}

\email{enricop@spotify.com}
\author{Nicola Tonellotto}
\orcid{0000-0002-7427-1001}
\affiliation{%
\institution{Pisa University \includegraphics[height=2ex]{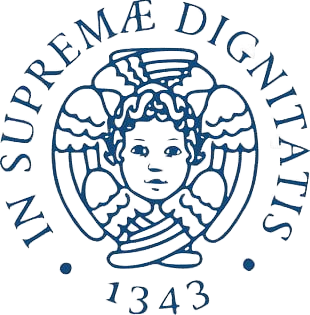}}
\city{Pisa}
\country{Italy}
}
\email{nicola.tonellotto@unipi.it}
\author{Fabrizio Silvestri}
\orcid{0000-0001-7669-9055}
\affiliation{%
\institution{Sapienza University\includegraphics[height=2ex]{images/logo2.png}, CNR \includegraphics[height=2ex]{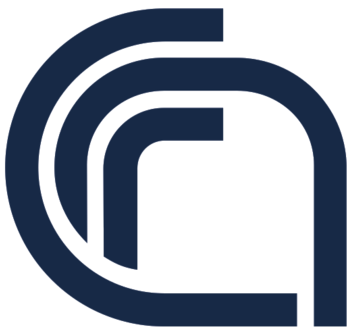}}
\city{Rome}
\country{Italy}}
\email{fsilvestri@diag.uniroma1.it}

\renewcommand{\shortauthors}{Bacciu, et al., 2024}

\begin{abstract}
Query recommendation systems are ubiquitous in modern search engines, assisting users in producing effective queries to meet their information needs.
However, many of these systems require a large amount of data to produce good recommendations, such as a large collection of documents to index and query logs. In particular, query logs and user data are not available in cold start scenarios. Query logs are expensive to collect and maintain and require complex and time-consuming cascading pipelines for creating, combining, and ranking recommendations.
To address these issues, we frame the query recommendation problem as a generative task, proposing a novel approach called Generative Query Recommendation (GQR).
GQR uses a Large Language Model (LLM) as its foundation and does not require to be trained or fine-tuned to tackle the query recommendation problem.
We design a prompt that enables the LLM to understand the specific recommendation task, even using a single example.
We then improved our system by proposing a version that exploits query logs called Retriever-Augmented GQR (RA-GQR).
RA-GQR can dynamically compose its prompt by retrieving similar queries from query logs.
GQR approaches reuses a pre-existing neural architecture resulting in a simpler and more ready-to-market approach, even in a cold start scenario. 
Our proposed GQR obtains state-of-the-art performance in terms of NDCG@10 and clarity score against two commercial search engines and the previous state-of-the-art approach on the Robust04 and ClueWeb09B collections, improving on average the NDCG@10 performance up to ${\sim}4\%$ on Robust04 and ClueWeb09B w.r.t the previous state-of-the-art competitor.
While the RA-GQR further improve the NDCG@10 obtaining an increase of ${\sim}11\%$, ${\sim}6\%$ on Robust04 and ClueWeb09B w.r.t the previous state-of-the-art competitor.
Furthermore, our system obtained ${\sim}59\%$ of user preferences in a blind user study, proving that our method produces the most engaging queries.
\end{abstract}
\maketitle


\section{Introduction}\label{sec:intro}


``Related Searches'' is a fundamental module of a Search Engines Result Page (SERP).
By recommending new queries related to the query submitted by a user, the related searches module guides the users toward their information needs by suggesting more focused and refined queries, called \textit{query recommendations}.
Typically, a ``Related Searches'' module is implemented by a \textit{query recommendation system.} 
To exemplify why related searches are important, consider a user trying to get information about the company \textit{Corporate Ltd.} the user enters the query ``\textsf{Corporate}'' and, by assuming this is a navigational query, the search engine's first result is the \textit{Corporate Ltd}'s website. 
This is not what the user was looking for, though, and the Related Searches component offers recommendations for queries such as ``\textsf{Corporate products}'', ``\textsf{Corporate job openings}'', and so on. Without needing to reformulate the query manually, the user is only ``one tap away'' from the SERP containing the information it was looking for.
%
However, from a practical point of view, building or improving a query recommendation system is challenging ~\cite{silvestri2009mining} as it involves three distinct requirements: past user-system interactions stored in \emph{query logs}, a machine-learned model trained on them, and a serving infrastructure to support the real-time recommendation in response to user queries. 


%
In contrast to most previous query recommendation solutions, here we show that it is possible to develop an effective ``Related Searches'' module \textit{\textbf{without}} the need for a system-specific machine-learned model trained on it. We present a novel way of conceiving a query recommendation system by using the great expressiveness and language generation power of Large Language Models (LLMs), such as OpenAI's GPT3~\cite{brown2020language}. 
%
An earlier attempt to build a query recommendation system without query logs was by Bhatia et al.~\cite{bhatia2011query} in 2011. 
However, when that research was done, the tools available to ``generate'' text were limited. In a nutshell, their approach was based on finding key phrases in a document collection whose prefixes were related to, or matching, the user-submitted query. 
However, this purely statistical word-based approach does not leverage complex correlations found in the language (which gives rise to learned semantics), and the language generation capabilities are also limited; these issues result in poor performance compared to modern query recommendation systems.

\begin{figure*}
    \includegraphics[width=0.9\textwidth]{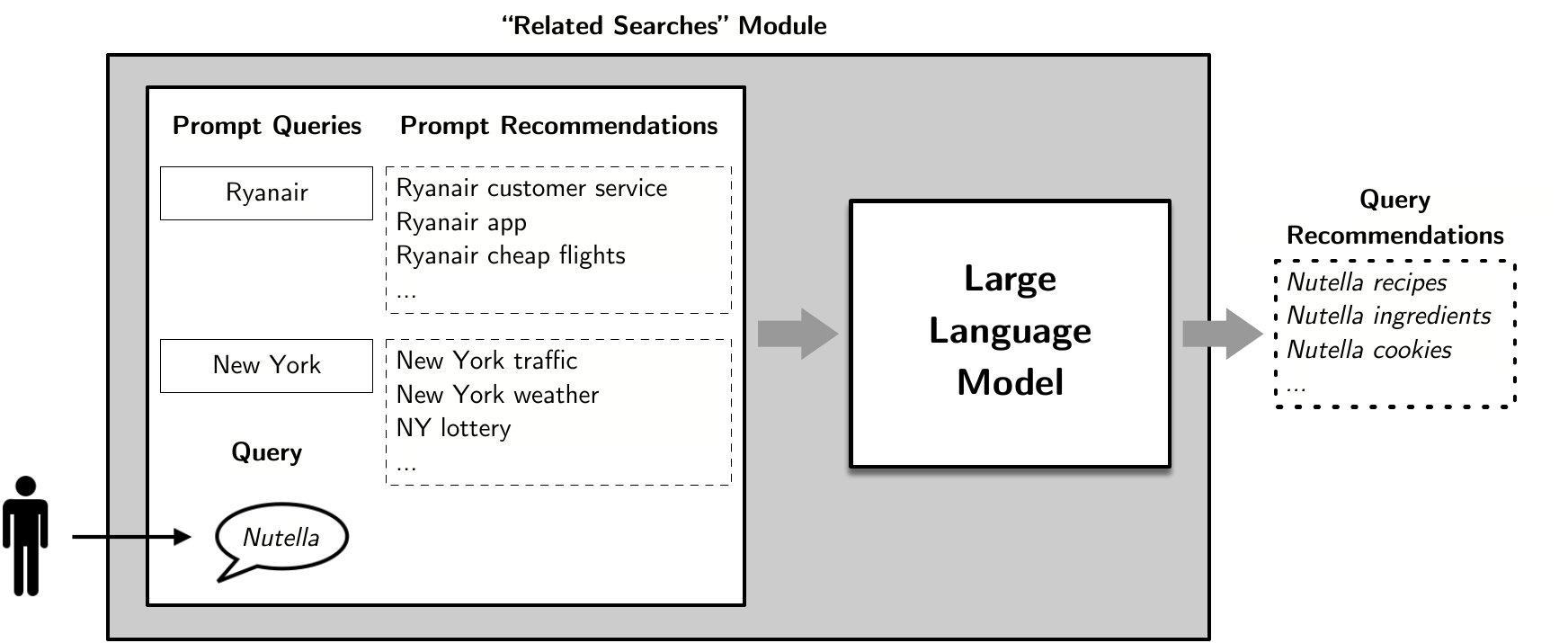}
    \caption{An example of prompt-based query recommendations for the query ``Nutella''. The Large Language Model is prompted with two queries ``Ryanair'' and ``New York'', and their relative recommendations, and a user query, and corresponding query recommendations are generated}
    \label{fig:gpt3example}
\end{figure*}

In this paper, we take a completely different approach inspired by the success of prompt-based zero-shot learning~\cite{10.1145/3560815}. We prompt an LLM, in our case, OpenAI's GPT3, with examples of queries and respective recommendations, and we let the LLM generate recommendations related to a given new query. An example is shown in Figure~\ref{fig:gpt3example}. The LLM is prompted with two queries, namely \textsf{Ryanair} and \textsf{New York}, together with their recommendations, and a last query, \textsf{Nutella}, for which the LLM generates recommendations.
Essentially, in that example, we build a prompt composed of two pairs of query recommendations, followed by the user query and a ``\textsf{recommendations}'' token.
In doing so, we ``instruct'' the LLM on what is the structure and the semantics of the continuation to the last token, and the system is able to find the relevant recommendations even if the category for the query is different from the two example queries in the prompt.
This is the essence of prompt-based approaches to zero-shot learning, providing an ``in-context'' set of examples that will drive the LLM generation phase toward solving a task. Dai et al.~\cite{dai2022can} recently postulated that prompt-based approaches are similar to meta-learning applied to a specific task. Prompts are comparable to support sets in meta-learners and act as a guide for generating the rest of the sentence, in our case, query recommendations.

From a time-to-market perspective, traditional query recommendation systems based on mining query logs~\cite{jones2006generating} suffer from the cold-start problem, as they need to collect and maintain large amounts of user data and query logs to estimate a meaningful correlation between co-observed queries in search sessions. This is far from ideal for newly launched search engines, but also for more established search engines that start operating in new regions or markets where the query logs may not be representative of the user behavior.
Our approach can greatly speed up the bootstrapping of a query recommendation system, as we only require a pre-trained LLM and a handful of hand-curated prompts, which manual annotators and domain experts can easily create.
These are the major benefits of using a prompt-based LLM approach for implementing a ``Related Searches'' component:

\begin{itemize}
\item There is no need to build a specific data structure and model for the query recommendation service. One can use an available LLM and apply the prompt we devise in this research.
\item Recommendations are generated without using any user information, past queries, or past interactions, except the current query. 
This greatly reduces the time-to-market of query recommendation systems and allows us to tackle the cold-start issue effectively. 
\item Generating recommendations for previously unseen and long-tail queries is a challenging task in traditional query recommender systems~\cite{bonchi2012efficient, sordoni2015hierarchical}. Using our technique, this task becomes trivial, and it does not require any specifically designed mechanisms to manage tail queries.
\end{itemize}

Query logs remain an important source of knowledge, for that reason, we propose an advanced version of GQR called Retrieval Augmented GQR (RA-GQR) able to leverage query logs to build automatically its prompt.
This version first retrieves similar users' written queries from a query log and then compose dynamically the GQR prompt to generate better recommendations. In that way the examples in the prompt are tailored within the topic of the current user query.

We introduce the Generative Query Recommendation (GQR) system. GQR uses an LLM as its foundation and leverages the prompting abilities of the LLM to understand the recommendation task through a few examples provided in the prompt (retrieved or handcrafted). Our preliminary experiments on two publicly available test collections show that GQR and RA-GQR outperform the other tested systems on average and obtain stable performance along query recommendations in the various ranks. For example, our GQR, exploiting the GPT-3 LLM, outperforms other query recommendation systems obtaining an NDCG@10 improvement of at least ${\sim}4\%$ in Robust04 and ClueWeb09B.
With the aid of 12 annotators, we conduct a blind user study that shows that GQR produces the most engaging and relevant query recommendations, with ${\sim}59\%$ preferences.

\section{Related Work}\label{sec:related-work}
In this section, we discuss the related work on query recommendations (Section~\ref{ssec:qs}), LLMs (Section~\ref{ssec:llm}), and prompting in Information Retrieval (IR) (Section~\ref{ssec:pir}).

\subsection{Query Recommendations}\label{ssec:qs}
Query recommendations are ubiquitous in modern search engines, assisting users in formulating and re-formulating effective queries and supporting exploratory searches. The query recommendation problem is generally modelled in two alternative ways. The first approach is based on the idea that the query recommendation provider is able to guide the user toward high-quality queries by looking into which queries have statistically led to successful search sessions in a set of query logs. For instance, the notion of query shortcut~\cite{baraglia2009search} corresponds to creating a link between the current user query and a successful query that is highly correlated with the user query, typically appearing at the end of a successful search session~\cite{10.1145/1935826.1935875}. This line of work is generally quite effective as it directly optimizes for successful searches, but it requires a way to determine whether a search session was successful, which is a non-trivial problem for public datasets, where metrics such as the stream or dwell time on a document are typically unavailable. Hence, many works in the literature have relied on the simplifying assumption that, within a search session, users optimize their queries until they find what they are looking for, making query reformulations within a search session a natural candidate for query recommendation. As a consequence, several models have relied on mining query-query correlations from query logs, either through more traditional statistical approaches such as Log-Likelihood Ratio~\cite{jones2006generating}, graph-based approaches~\cite{huang2003relevant, boldi2009query, craswell2007random, beeferman2000agglomerative} or using modern language technologies that can generalize to new queries and effectively model sequences such as RNNs~\cite{wu2018query} or transformers~\cite{mustar2021study}. Nevertheless, these approaches still require access to a set of high-quality query logs, which are expensive to collect, maintain, and might not be available or be representative in a cold-start situation, e.g. launch in a new country, for a new product, etc. To deal with cold-start issues, some approaches for query recommendations in the absence of query logs have been proposed in the past. For example,  Bhatia et al.~\cite{bhatia2011query} build a (pre-neural) language model, that is based on a database of sentences, such as those contained in the document collection. This language model is obtained by extracting the n-grams (in the order of unigram, bigram and trigram) from the collection. Then, given a user query, their approach relies on the last word of the user's query to find the most probable subsequent term to obtain a phrase completion with maximum likelihood estimation. However, these collection-extractive approaches are based on finding relevant key-phrases in a collection of documents and do not take advantage of the outstanding generative capabilities of Large Language Models, which have shown to be extremely effective in IR tasks such as query generation (see Section~\ref{ssec:pir}). To the best of our knowledge, this is the first approach to generate query recommendations that only requires a pre-trained neural network, without any access to user data or documents.

\subsection{Large Language Models}\label{ssec:llm}

Pre-trained Language Models (LMs) such as BERT~\cite{devlin2018bert} and T5~\cite{t5}  are nowadays the main building block of the major Natural Language Processing (NLP) architectures.
After being fine-tuned, these LMs demonstrate impressive capabilities in a variety of tasks such as Question Answering, Machine Translations, Semantic Parsing, and so on~\cite{lin2020pretrained}.
Recent advances in the field of language modelling have led to the creation of pre-trained Large Language Models (LLMs) such as GPT~\cite{brown2020language}, which are LMs with a huge number of parameters, trained on large amounts of data. While LMs require task-specific fine-tuning to update their parameters to correctly address a specific downstream task, LLMs are able to perform in-context learning, i.e., the ability to address a specific downstream task by learning from just a few examples in the task demonstration context~\cite{wei2022chain,wei2022emergent}. These few examples are usually written in natural language templates, or \textit{prompts}. In-context learning concatenates a few examples of the demonstration context together with a question, which is fed into the LLM to generate an answer. In doing so, the LLM does not require supervised fine-tuning, and directly performs inference using the provided prompt.
For example, given a LLMs and an english sentence, we can condition the model's output by asking to address different tasks such as \textit{Translate the utterance in German} or \textit{Paraphrase this utterance}, and we can also combine them like \textit{Translate and paraphrase this sentence in German}.

Many LLM have been developed and trained, such as GPT3~\cite{brown2020language}, BLOOM~\cite{scao2022bloom}), and OPT~\cite{zhang2022opt}). Despite their capabilities, the performance of LLMs is sensitive to the prompting template~\cite{zhao2021calibrate}. For our solution we propose a task-specific prompt to generate query recommendations, and provides the first evaluation of a LLM's in-context learning for the query recommendations generation task.

\subsection{Prompting in Information Retrieval}\label{ssec:pir}

Nogueira et al.~\cite{monot5} have been among the first to propose a prompt-based approach for relevance ranking using a LM~\cite{t5}. Given a query and a document, their texts are concatenated with task-specific tokens, and the task was to predict if the document would be relevant for the given query. However, the proposed template has been used to generate training samples to fine-tune the LM on the relevance classification task. 
Sachan et al.~\cite{sachan2022improving} demonstrate how prompting can be used for re-ranking documents without fine-tuning, i.e., in a zero-shot setting. Their re-ranker scores documents with a zero-shot LM by computing the probability of the input query conditioned on a retrieved document. 
InPars~\cite{inpars,inpars2} and Promptagator~\cite{dai2022promptagator} demonstrated how prompting and LLM can be used to generate synthetic training datasets for IR tasks.
Both models create a prompt composed of a concatenation of few query-relevant document pairs followed by a document, and the goal is to generate a query for which the final document is relevant.
This prompt is then used with the GPT3~\cite{inpars,dai2022promptagator}, GTP-J~\cite{wang2021gpt}, or FLAN~\cite{wei2021finetuned} LLMs to generate new synthetic query-document pairs. Recently, prompt-based IR pushed as far as replacing the classic index-retrieval pipeline with a single large language model that directly generates relevant document IDs given a user query~\cite{tay2022transformer}.

Inspired by the successes of these works that use prompting to generate queries, we focus on prompting the query recommendation generation problem.

\section{Proposed Approach}\label{sec:proposed-approach}
We reframe the query recommendation problem as a generative task introducing Generative Query Recommendation (GQR). 
GQR uses a pre-trained sequence-to-sequence LLM as its foundation, and it leverages the prompting abilities of the LLM to understand the recommendation task through a few examples provided in the prompt.
Indeed, our methodology consists in building a proper prompt composed of pairs of queries and a list of corresponding query recommendations to instruct an LLM to generate recommendations for a new, previously unseen user query.

More formally, given a query input text $q$ and a candidate recommendations set $\mathcal{L} = \{r_1, r_2, \ldots\}$, i.e., a set of free-text recommendations, a pre-trained LLM $\mathcal{M}$ takes the top $k$ candidate recommendations in $\mathcal{L}$ with the highest probabilities of the language model $\mathcal{M}$, conditioned on $n$ examples $\mathcal{C} = \big\{(q_1', \ell_1'), (q_2', \ell_2'), \ldots, (q_n', \ell_n')\big\}$, where $(q_i', \ell_i')$ is a query-list of recommendations pair whose query is different from the query input $q$.

Since the recommendation space $\mathcal{L}$ is potentially very large, we do not explicitly compute and store it, but we leverage the text generation capabilities of $\mathcal{M}$ to synthesize recommendations. Our proposed GQR system generates the highest scoring $k$ recommendations $\hat{\ell} = \{\hat{r}_1, \ldots, \hat{r}_k\} \subset \mathcal{L}$ by selecting the $k$ elements of $\mathcal{L}$ with the highest generation probability:
\begin{equation}
    \hat{\ell} = k\text{-arg}\max_{r_j \in \mathcal{L}} P_{\mathcal{M}}(r_j|q,\mathcal{C}).
\end{equation}

In practice, we use a pre-defined template, or prompt, to format the examples and pre-append them to the input query. Let $T(\cdot, \cdot)$ the formatting function of an input example, i.e.,
\begin{equation}
    T(q,\ell) = \mathsf{query}\;q\; \mathsf{recommendations}\;\ell
\end{equation}
The context $\mathcal{C}$ and the input query $q$ are provided as input to the the LLM $\mathcal{M}$ according to the following prompt:
\begin{equation}
    T(q_1', \ell_1')\;\;T(q_2', \ell_2')\;\;\cdots\;\;T(q_n', \ell_n')\;\;T(q,\_)
\end{equation}

Then, to generate the query recommendation, we provide the input build according to this prompt to the LLM, and we let the LLM generate the recommendations, as shown in Figure~\ref{fig:gpt3example}.
Hence, an example of our prompt is 
\begin{lstlisting}
query: $q_1$
recommendations: $\ell_1$
$\vdots$
query: $q_n$
recommendations: $\ell_n$
query: $q'$
recommendations:
\end{lstlisting}
Then, our LLM will condition its output using the prompt. It will continue to complete the input text adding the recommendations for the query $q'$.

\begin{figure*}
    \includegraphics[width=\textwidth]{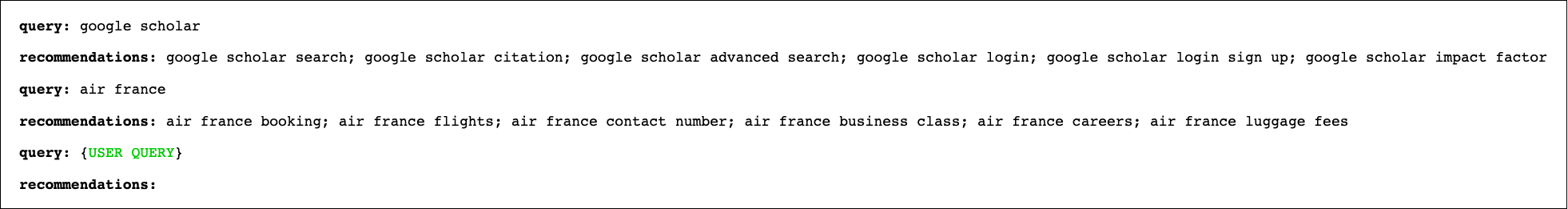}
    \caption{Instance of one of our prompts with two examples. As can be noticed from this Figure, we concatenated the user query with the prompt and then we let the LLM to continue the generation of the query recommendations.}
    \label{fig:prompt}
\end{figure*}

To get the most general prompt possible, we use different examples that cover several topics, as shown in Figure \ref{fig:prompt}. The examples composing our prompt include queries related to entities of different natures, such as company names, famous person names, product names, historical events, etc. For each prompt, we built hand-crafted recommendations, which will be released upon acceptance.

Prompting brings several advantages in query recommendation generation, considerably easing the different constraints required in the current query recommendation systems.
GQR consists of a prompt and a general-purpose LLM, that, by using an in-context learning strategy, is able to generate query recommendations in a single forward pass, and it does not need additional \textit{ad hoc} systems to produce, rank, and combine query recommendations. The system's overall inductive biases that affect the final output are easily controllable through the human-created prompt in plain language. 
Further, GQR exploits the large pre-training knowledge of the LLM to retrieve the information needed to generate the recommendations, removing the need to index large collections of documents to extract query recommendations.

By re-using the pre-trained knowledge of the LLM, we obtain an approach that does need specific training (or fine-tuning) to produce query recommendations.
In that way, we discontinue the use of users' data, such as query logs, because no data is involved in our query suggestion pipeline, resulting in an approach that is robust to cold-start. Indeed, the LLM is typically pre-trained on an extremely large and diverse set of texts, allowing it to distill both language and factual knowledge into a single model. It, therefore, equips our approach with an open, unconstrained source of recommendations in contrast to being bound to a finite set of recommendations extracted from a query log. As for the language aspect, our approach leverages all the common advantages of pre-trained transformer architectures \citep{devlin2018bert}, such as Word Pieces embeddings \citep{wu2016google}, making the model more robust to typing mistakes \citep{zhang2020spelling} and able to handle different word forms effectively, and able to understand the context and semantics of the query.

To sum up, this approach completely eradicates the cold-start issues, it demonstrates that there is no need for large collections of documents, to generate query recommendations, and the need to construct a query-specific architecture, allowing the reuse of existing models, thus promoting a greener approach \cite{patterson2021carbon}.

\section{Experimental Setup}\label{sec:expsetup}
In this section, we describe our experimental setup, introducing our research questions (RQs) and the datasets, the competing models, the hardware, and the evaluation metrics used to answer our RQs.

\subsection{Research Questions}\label{sec:research-questions}

In our experiments, we focus on answering the following research questions
\begin{itemize}
    \item \textbf{RQ1:} Can our proposed GQR system generate relevant and useful query recommendations compared with existing query recommendation systems?
    \item \textbf{RQ2:} Are the queries generated by our GQR system more engaging for users than those generated by other systems?
    \item \textbf{RQ3:} Does our GQR system generate recommendations for long tail, i.e., rare,  queries?
    \item \textbf{RQ4:} Are query logs still bring value in generative query recommendation?
\end{itemize}

\subsection{Datasets}\label{sec:datasets}
We use three datasets, TREC Robust 2004 Disk 4-5 \cite{Voorhees1996Disks45, Voorhees2004Robust, Huston2014ACO}.
 with 250 queries and 528,155 documents, ClueWeb09B \cite{Clarke2009TrecWeb} with 200 queries and 50,220,423 documents and a subset of queries from the AOL dataset \cite{pass2006picture} , specifically 192 queries randomly sampled to perform a user study and the first 200 queries extracted from the query log's long tail distribution, i.e., queries appearing just once in the log. We will release the datasets upon acceptance.

\subsection{Tested Models}\label{sec:models}
We compare our proposed GQR approaches with two query recommendation systems implemented by publicly available Web search engines, referred to as \textit{\textit{System~1}} and \textit{\textit{System~2}} to maintain anonymity. To compute their performance, we collect the recommendations they produce by scraping their outputs in response to our test queries.
Moreover, we compare GQR against the most similar query recommendation system available in the research literature, which means a system not exploiting query logs for producing query recommendations~\cite{bhatia2011query}.
Given that their implementation cannot be found, we reproduce their work in our experiments.
We conducted the experiments of Bhatia et al. 2011 \cite{bhatia2011query} only on Robust 2004 because computing $n$-grams on a large collection like ClueWeb09B would be too demanding on memory. \footnote{Indeed, by a simple back-of-the-envelope calculation, the approach of Bhathia et al. would require more than 1TB of RAM to store just the index.
By observing the experimental results we shall present in the next sections, we decided that adding this experiment would not have added any additional insights.}

We implement our GQR system using two different LLMs: GPT-3~\cite{brown2020language} using the implementation of \textit{text-davinci-003} model with 175B parameters accessed through the OpenAI APIs\footnote{\url{https://platform.openai.com/playground}}, and Bloom~\cite{scao2022bloom} with 176B parameters, using the Huggingface's APIs\footnote{\url{https://huggingface.co/bigscience/bloom}}. Both GQR systems exploit a prompt composed of $10$ examples by default.

We implement RA-GQR using our best-performing model, GPT-3. 
RA-GQR leverages past user queries from Lucchese et al. (2013) ~\cite{lucchese2013modeling}.  Instead of relying on generic prompts, RA-GQR dynamically builds its prompts by incorporating real user queries similar to the current one. This approach aims to provide GQR with relevant in-context learning examples. To achieve this, RA-GQR first embeds the queries using a pre-trained model, \texttt{msmarco-roberta-base-v2}. Then, it employs FAISS \cite{douze2024faiss} to efficiently retrieve past user queries that share the same topic as the current query.
RA-GQR builds upon the concept of Retrieval Augmented Generation (RAG) introduced by Lewis et al. \cite{lewis2020retrieval}.

For all the GQR implementations, we use the default hyper-parameters: temperature of $0.7$, with a maximum of $256$ input tokens for both models.
To maintain consistency with the outputs of \textit{System~1} and \textit{System~2}, which typically generate six related searches, we decided to evaluate the top six query recommendations generated by GPT3 and Bloom.



\subsection{Evaluation Protocols}
To assess the performance of query recommendation systems, we exploit two different evaluation protocols.
Given a query $q$ and a set of $k$ query recommendations  $r_1, r_2, ..., r_k$ generated by a given query recommendation system, our first protocol, named \textit{Substitution}, evaluates the effectiveness of each recommendation independently, i.e., the original query is replaced with one of the generated recommendations:
\begin{equation}
    \text{Substitution}(q,i) \equiv r_i.
\end{equation}
In doing so, we simulate a user selecting a single query recommendation to be used in place of the submitted query.

However, this protocol does not take into account the query recommendation capabilities of a system as a whole. Hence we introduce
our second protocol, named \textit{Concat},
consisting of an iterative concatenation of the original query with the first $i$ query recommendations:
\begin{equation}
    \text{Concat}(q,i) \equiv q \oplus r_1 \oplus \cdots \oplus r_i.
\end{equation}
The goal of this protocol is to assess the increase in performance that a user might get when the search results are (possibly) enriched by the additional information coming from the results retrieved thanks to the considered set of query recommendations.

\subsection{Performance Metrics}

\hspace{1em}\textit{Simplified Clarity Score.}
A common metric used in query recommendations is the Clarity Score (CS)~\cite{cronen2002predicting}, used to measure the specificity or ambiguity of a query. We exploit the Simplified Clarity Score (SCS) implementation~\cite{he2004inferring}, which is less time-consuming when compared to the complexity of Clarity Score.

SCS relies on calculating a relevance model based on the query and compares it to the relevance model generated by the entire corpus using Kullback–Leibler Divergence:
\begin{equation}
    SCS(q) = \sum_{w \in q} p(w|q) \log_2 \frac{p(w|q)}{p(w|C)},
\end{equation}
where $p(w|q)$ is the statistical language model built over the query, while $p(w|C)$ is the statistical language model built over the document collection
The quantity $p(w|q)$ is estimated with the maximum likelihood of the query model of the word $w$ in query $q$ and it is computed as the ratio between the number of occurrences of the word $w$ in the query and the query length.
The quantity $p(w|C)$ is computed as the ratio between the number of occurrences of the word $w$ in the collection $C$ and the total number of words in $C$.
In other words, a high SCS indicates a well-formed query with little ambiguity that leads to retrieving a highly coherent set of documents. It has been shown to correlate well with relevance judgments and it is widely used as a query performance predictor in absence of explicit relevance judgments~\cite{cronen2002predicting}.

\textit{Retrieval effectiveness.}
As mentioned in Section~\ref{sec:intro}, query recommendations are a useful tool to help users in producing better queries to retrieve the documents that satisfy their information needs.
So, to assess the retrieval effectiveness of the generated query recommendations, we test their effectiveness to retrieve relevant documents concerning the original query. Our reference effectiveness metric is the Normalized Discounted Cumulative Gain at cutoff 10 (NDCG@10). To perform our experiments, we used the BM25 ranking function from the PyTerrier library~\cite{macdonald2021pyterrier}.

All statistical significance differences are computed with the paired t-test ($p < 0.01$) with the Holm-Bonferroni multiple testing correction.

\subsection{User study}\label{sec:user-study}
The main goal of a query recommender system should be that of producing helpful recommendations, not only effective from a retrieval point of view recommendations.
Therefore, we aim to determine which system, amongst the ones we tested, presents the most engaging query recommendations from a user's perspective.
We engage twelve professional annotators with the recommendations generated by \textit{System~1}, \textit{System~2}, and GQR (GPT-3). We divide annotators into three groups, i.e., four annotators per group. We also randomly partition the 192 queries randomly sampled from the AOL query log into the above-mentioned three groups, ending up with 64 queries per group.

The user study is conducted as follows: each system is anonymously identified and shuffled for each query to eliminate bias.
In the user study, our input queries always get at least six recommendations from each system. The order of the recommendations shown to the annotators is not taken into account.
We ask the annotators to choose the most helpful and engaging set of recommended queries that do not repeat the same information.
As an example of our guideline recommendations, if one of the systems under study provides recommendations such as "Ryanair support, Ryanair contact, Ryanair customer service" w.r.t. the input query "Ryanair", that system is not helpful in satisfying the user's information need because it is just repeating the same recommendation.
A better system would provide diverse recommendations, such as "Ryanair careers, Ryanair history, Ryanair cheap destinations."



\section{Results and Analysis} \label{sec:results-analysis}

In this section, we present the results to answer our posed five
research questions (Section \ref{sec:research-questions}).



\textbf{RQ1: Can our proposed GQR system without query-logs generate relevant and useful query recommendations compared with existing query recommendation systems?}

In Tables \ref{tab:scs-subs} and \ref{tab:ndcg-subs} we report the simplified clarity score and the NDCG@10 metrics, respectively, for the tested query recommendation systems, according to the \textit{Substitution} evaluation protocol. For each system, collection, and metric, we report the minimum and maximum metric values across the six generated recommendations, as well as their average values and the standard deviations.
According to the SCS metric, our GQR systems get the highest values with both the worst and the best-performing recommendations on the Robust04 dataset, outperformed only by \textit{System~2}'s best recommendation on the ClueWeb09B dataset, but by a small margin. However, on average, our proposed GQR (GPT-3) system outperforms all other methods, including commercial systems. Moreover, the corresponding standard deviation values are the smallest, meaning that all the generated query recommendations obtain a similar, high SCS score.
According to the NDCG@10 metric, our proposed GQR (GPT-3) system outperforms all other systems. With respect to the best competitor, namely \textit{System~2}, our experiments show that GQR (GPT-3), on average, has relative improvements of +23.86\% and a +26.63\% in NDCG@10 on Robust04 and ClueWeb09B, respectively. As with the SCS metric, the standard deviation of the NDCG@10 scores across the six generated recommendations is small, so the recommendations get an NDCG@10 score concentrated around the average value.

\begin{table}[ht!]
\centering
\begin{tabular}{l c c c }
\toprule
\textbf{Model} & \textbf{Min} & \textbf{Max} & \textbf{Avg $\pm$ Std}\\
\midrule
\multicolumn{4}{c}{\textbf{Robust04}}\\
\midrule
\textit{System~1} (a)              &           9.21 &          10.17 &          9.80 $\pm$ 0.35 \\
\textit{System~2} (b)              &           9.35 &          10.56 &          9.82 $\pm$ 0.39 \\
Bhatia~\cite{bhatia2011query} (c)  &           7.64 &           9.20 &          8.28 $\pm$ 0.53 \\
GQR (Bloom) (d)                    &           7.50 &          11.00 &          8.84 $\pm$ 1.66 \\
GQR (GPT-3) (e)                    &          10.54 &          10.77 &          10.65 $\pm$ 0.08\\
RA-GQR (GPT-3)                     & \textbf{16.71}$^{abcde}$ &    \textbf{17.10}$^{abcde}$ & \textbf{16.98 $\pm$ 0.19}$^{abcde}$\\
\midrule
\multicolumn{4}{c}{\textbf{ClueWeb09B}}\\
\midrule
\textit{System~1} (a)              &           9.80 &          10.80 &         10.37 $\pm$ 0.32  \\
\textit{System~2} (b)              &          10.49 &           11.31 &         10.87 $\pm$ 0.30  \\
Bhatia~\cite{bhatia2011query} (c)  &              - &              - &                        -  \\
GQR (Bloom) (d)                    &           7.68 &          11.20 &          9.84 $\pm$ 1.19  \\
GQR (GPT-3) (e)                       & 10.94          &   11.22        &          11.12 $\pm$ 0.10 \\
RA-GQR (GPT-3)                      & \textbf{19.42}$^{acde}$ &   \textbf{19.76}$^{acde}$ & \textbf{19.53 $\pm$ 0.20}$^{abcde}$ \\

\bottomrule
\end{tabular}
\caption{SCS for the \textit{Substitution} protocol for each system on Robust04 and ClueWeb09B. The best values across all systems are boldfaced. The letters indicate a statistically significant difference w.r.t. GQR (GPT-3).}
\label{tab:scs-subs}
\end{table}

\begin{table}[ht!]
\centering
\begin{tabular}{l c c c }
\toprule
\textbf{Model} & \textbf{Min} & \textbf{Max} & \textbf{Avg $\pm$ Std}\\
\midrule
\multicolumn{4}{c}{\textbf{Robust04}}\\
\midrule
\textit{System~1} (a)              & 0.2038 & 0.2638 & 0.2377 $\pm$ 0.0211 \\
\textit{System~2} (b)              & 0.2546 & 0.3739 & 0.3102 $\pm$ 0.0478 \\
Bhatia~\cite{bhatia2011query} (c)  & 0.2388 & 0.2640 & 0.2566 $\pm$ 0.0108 \\
GQR (Bloom) (d)                    & 0.1743 & 0.3655 & 0.2628 $\pm$ 0.0714 \\
GQR (GPT-3) (e)                       & 0.3727 & 0.3947 & 0.3842 $\pm$ 0.0092 \\
RA-GQR (GPT-3)                     & \textbf{0.4197}$^{abde}$ & \textbf{0.4476}$^{abcde}$ & \textbf{0.4339 $\pm$ 0.008}$^{abcd}$ \\
\midrule
\multicolumn{4}{c}{\textbf{ClueWeb09B}}\\
\midrule
\textit{System~1} (a)              & 0.0940 & 0.1129 & 0.1056 $\pm$ 0.0082 \\
\textit{System~2} (b)              & 0.0946 & 0.1294 & 0.1108 $\pm$ 0.0164 \\
Bhatia~\cite{bhatia2011query} (c)  & - & - & -  \\
GQR (Bloom) (d)                    & 0.0468 & 0.1144 & 0.0802 $\pm$ 0.0268 \\
GQR (GPT-3) (e)                      &  0.1280 & 0.1659 & 0.1403 $\pm$ 0.0146 \\
RA-GQR (GPT-3)                        & \textbf{0.1597}$^{abde}$ & \textbf{0.1832}$^{abde}$ & \textbf{0.1708 $\pm$ 0.10}$^{abde}$ \\

\bottomrule
\end{tabular}
\caption{NDCG@10 for the \textit{Substitution} protocol for each system on Robust04 and ClueWeb09B. The best values across all systems are boldfaced. The letters indicate a statistically significant difference w.r.t. GQR (GPT-3).}
\label{tab:ndcg-subs}
\end{table}

In Tables \ref{tab:scs-conc} and \ref{tab:ndcg-conc}, we report the clarity score and the NDCG@10 metrics, respectively, for the tested query recommendation systems, according to the \textit{Concat} evaluation protocol. This protocol aims to assess the performance improvements that a user query may get when the generated query recommendations are used as additional information sources together with the query. In doing so, the query recommendations are expected to better specify the information need of the user expressed as a text query.
In doing so, it is important to take into account the number of query recommendations concatenated with the user query, so we report the metric scores for increasing the number of recommendations concatenated, referred to as \textit{rank}
In both datasets, the concatenation of one or more query recommendations increases the SCS value, for all tested systems, as expected. In fact, according to \citet{cronen2002predicting}, the clarity score of a query increases if we add terms that reduce the query ambiguity. Since we are enriching the user queries with recommendations, these increases confirm that the generated recommendations do not increase the ambiguity of the queries. Among the tested systems, our proposed GQR (GPT-3) outperforms all other systems across all ranks and datasets.
Similar results are obtained w.r.t. NDCG@10. In this case, the more recommendations we concatenate to the query, the higher the metric value. With six recommendations, the effectiveness performance increases by ${\sim}16.44\%$ on the Robust04 dataset and by ${\sim}22.58\%$ on the ClueWeb09 dataset by comparing with their single query recommendation with the highest NDCG@10.

Concerning RQ1, we can conclude that our proposed GQR (GPT-3) system is able to generate query recommendations without query logs that are, on average, less ambiguous than other systems and better or on par with commercial competitors.

\begin{table*}[ht!]
\centering
\begin{tabular}{l c c c c c c}
\toprule
\textbf{Model} & \textbf{rank 1} & \textbf{rank 2} & \textbf{rank 3} & \textbf{rank 4} & \textbf{rank 5} & \textbf{rank 6}\\
\midrule
\multicolumn{7}{c}{\textbf{Robust04}}\\
\midrule
\textit{System~1} (a)             & 16.29 & 20.00 & 23.32 & 26.55 & 29.18 & 31.43 \\
\textit{System~2} (b)             & 19.35 & 24.81 & 31.10 & 36.63 & 42.84 & 47.88 \\
Bhatia~\cite{bhatia2011query} (c) & 18.21 & 23.89  & 29.52  & 34.49  & 39.56 & 44.28 \\
GQR (Bloom) (d)                    & 19.13 & 25.41 & 31.08 & 36.12 & 40.48 & 44.20 \\
GQR (GPT-3) (e)                       & 20.00 & 27.69 & 34.98 & 42.23 & 49.47 & 56.67 \\
RA-GQR (GPT-3)                        & \textbf{27.23}$^{abcdef}$ & \textbf{40.69}$^{abcdef}$ & \textbf{54.51}$^{abcdef}$ & \textbf{68.09}$^{abcdef}$ & \textbf{81.98}$^{abcdef}$ & \textbf{95.48}$^{abcdef}$ \\
\midrule
\multicolumn{7}{c}{\textbf{ClueWeb09B}}\\
\midrule
\textit{System~1} (a) & 19.20 & 24.89 & 30.54 & 36.03 & 41.23 & 46.24 \\
\textit{System~2} (b) & 19.28 & 24.50 & 30.77 & 35.55 & 41.55 & 46.70 \\
Bhatia~\cite{bhatia2011query} (c) & -  & - & - & - & - & - \\
GQR (Bloom) (d) & 18.49 & 25.20 & 31.15 & 37.07 & 42.28 & 46.72 \\
GQR (GPT-3) (e) & 20.41 & 27.73 & 35.01 & 42.27 & 49.30 & 56.25 \\

RA-GQR (GPT-3) & \textbf{29.63}$^{abcde}$ & \textbf{46.43}$^{abcde}$ & \textbf{63.12}$^{abcde}$ & \textbf{79.32}$^{abcde}$ & \textbf{95.87}$^{abcde}$ & \textbf{112}$^{abcde}$ \\

\bottomrule
\end{tabular}
\caption{SCS for the \textit{Concat} protocol for each system on Robust04 and ClueWeb09B. The best values per rank across all systems are boldfaced. The letters denote a statistically significant difference w.r.t. GQR (GPT-3).}
\label{tab:scs-conc}
\end{table*}

\begin{table*}[ht!]
\centering
\begin{tabular}{l c c c c c c}
\toprule
\textbf{Model} & \textbf{rank 1} & \textbf{rank 2} & \textbf{rank 3} & \textbf{rank 4} & \textbf{rank 5} & \textbf{rank 6}\\
\midrule
\multicolumn{7}{c}{\textbf{Robust04}}\\
\midrule
\textit{System~1} (a)             & 0.4151 & 0.4270 & 0.4320 & 0.4373 & 0.4450 & 0.4417  \\
\textit{System~2} (b)             & 0.4236 & 0.4292 & 0.4361 & 0.4386 & 0.4433 & 0.4373  \\
Bhatia~\cite{bhatia2011query} (c) & 0.4044 & 0.4012 & 0.3956 & 0.3917 & 0.3900 & 0.3922  \\
GQR (Bloom) (d)                   & 0.4252 & 0.4254 & 0.4149 & 0.4136 & 0.4172 & 0.4203  \\
GQR (GPT-3) (e)                   & 0.4316 & 0.4429 & 0.4507 & 0.4543 & 0.4597 & 0.4596 \\
RA-GQR (GPT-3)             & \textbf{0.4553}$^{abcd}$ & \textbf{0.4690}$^{abcde}$ & \textbf{0.4788}$^{abcde}$ & \textbf{0.4869}$^{abcde}$ & \textbf{0.4876}$^{abcde}$ & \textbf{0.4889}$^{abcde}$ \\

\midrule
\multicolumn{7}{c}{\textbf{ClueWeb09B}}\\
\midrule
\textit{System~1} (a)             & 0.1598 & 0.1627 & 0.1832 & 0.1797 & 0.1912 & 0.1954  \\
\textit{System~2} (b)             & 0.1591 & 0.1429 & 0.1506 & 0.1528 & 0.1562 & 0.1645  \\
Bhatia~\cite{bhatia2011query} (c) & -  & - & - & - & - & - \\
GQR (Bloom) (d)                   & 0.1586 & 0.1626 & 0.1576 & 0.1544 & 0.1443 & 0.1427  \\
GQR (GPT-3) (e)                      & \textbf{0.1960} & 0.1891 & 0.1919 & 0.1939 & 0.2010 & 0.2030 \\
RA-GQR (GPT-3)                    & 0.1937$^{abcd}$ & \textbf{0.1951}$^{abcd}$ & \textbf{0.1985}$^{abcd}$ & \textbf{0.2003}$^{abcd}$ & \textbf{0.2078}$^{abcd}$ & \textbf{0.2058}$^{abcd}$ \\

\bottomrule
\end{tabular}
\caption{NDCG@10 for the \textit{Concat} protocol for each system on Robust04 and ClueWeb09B. The best values per rank across all systems are boldfaced. Our tests do not report statistically significant differences  w.r.t. GQR (GPT-3).} 
\label{tab:ndcg-conc}
\end{table*}

\textbf{RQ2: Are the queries generated by our GQR system more engaging for users than those generated by other systems?}\\
To answer this research question, we report the results of our user study in Table~\ref{tab:user-study}.
We exclude Bathia's approach because it needs to index a collection of documents which is not provided with the AOL queries.
The user study shows a clear preference for our GQR (GPT-3) system with respect to the other two commercial systems. All groups report a preference above $50\%$ of tested queries for the GQR (GPT-3) system, with an overall preference of ${\sim}59\%$. 
The second most preferred system is \textit{\textit{System~1}} with ${\sim}26\%$ of the preferences on average, and the last system is \textit{\textit{System~2}}, with ${\sim}15\%$ of preferencess on average.
Hence, on RQ2, we can conclude that, according to our user study, the recommendations generated by our GQR (GPT-3) system are more engaging than the recommendations generated by the two commercial competitors from the user's point of view.
It is worth noticing that AOL queries cover a wide variety of topics and since it is impossible to make assumptions about users' prior knowledge and their personal preferences, for that reason, subjectivity plays a crucial role.

\begin{table}[ht!]
\centering
\begin{adjustbox}{max width=\columnwidth}
\begin{tabular}{c c c c c}
\toprule
\textbf{Annotator ID} & \textbf{\textit{System~1}} & \textbf{\textit{System~2}} & \textbf{GQR (GPT-3)}\\
\midrule
\multicolumn{4}{c}{\textbf{Group 1}}\\
\midrule
Ann 1 & 20.31\% & 15.62\% & \textbf{64.06\%} \\
Ann 2 & 32.81\% & 15.63\% & \textbf{51.56\%} \\
Ann 3 & 31.25\% & 18.75\% & \textbf{50.00\%} \\
Ann 4 & 26.98\% & 11.11\% & \textbf{61.90\%} \\
Avg $\pm$ Std & 27.84\% $\pm$ 4.84\% & 15.28\% $\pm$ 2.72\% & \textbf{56.88\%} $\pm$ 6.17\% \\
\midrule
\multicolumn{4}{c}{\textbf{Group 2}}\\
\midrule
Ann 5 & 20.31\% & 23.44\% & \textbf{56.25\%} \\
Ann 6 & 39.06\% & 14.06\% & \textbf{46.88\%} \\
Ann 7 & 25.00\% & 12.50\% & \textbf{62.50\%} \\
Ann 8 & 31.25\% & 17.19\% & \textbf{51.56\%} \\
Avg $\pm$ Std & 28.91\% $\pm$ 7.03\% & 16.80\% $\pm$ 4.19\% & \textbf{54.30\%} $\pm$ 5.78\% \\
\midrule
\multicolumn{4}{c}{\textbf{Group 3}}\\
\midrule
Ann 9 & 17.19\% & 18.76\% & \textbf{64.06\%} \\
Ann 10 & 26.56\% & 9.38\% & \textbf{64.06\%} \\
Ann 11 & 18.75\% & 14.06\% & \textbf{67.19\%} \\
Ann 12 & 25.00\% & 4.69\% & \textbf{70.63\%} \\
Avg $\pm$ Std & 21.88\% $\pm$ 3.98\% & 11.72\% $\pm$ 5.24\% & \textbf{66.41\%} $\pm$ 2.71\%\\
\midrule
Overall & 26.21\% $\pm$ 6.26\% & 14.60\% $\pm$ 4.69\% & \textbf{59.19\% }$\pm$ 7.33\% & \\
\bottomrule
\end{tabular}
\end{adjustbox}
\caption{The annotator preferences (in percentage) in our user study, for each group and system under evaluation.}
\label{tab:user-study}
\end{table}
\textbf{RQ3: Does our GQR system generate recommendations for long tail, i.e., rare,  queries?}\\
During our experiments on Robust04 and ClueWeb09B, we observed that some systems fail to produce query recommendations for some queries, while our GQR (GPT-3) system is always successful in generating at least 1 suggestion for each query.
To further investigate on this topic, we extracted 200 rare queries from the tail distribution of the AOL query log, thus considering the queries with a frequency equal to 1-2 in the query log.
In Table \ref{tab:long-tail} we report, in percentage, the number of times our tested systems have been able to generate at least one and all six recommendations for the queries Robust04, ClueWeb09B and the 200 AOL tail queries datasets. We also report the average number of query recommendations each system successfully generated for each dataset.
For the Robust04 and ClueWbe09B datasets, all competitors are able to generate more than 5 recommendations on average. On average, all systems generate at least one recommendation for more than $90\%$ of the Robust04 queries, and for more than $98\%$ on ClueWeb09B. \textit{System~1} and \textit{System~2} are not able to generate at least one recommendation $9\%$ and $17\%$ of the times on AOL tail queries, respectively. Our GQR (Bloom) system generates at least one query recommendation $99\%$ of the times, but it is the worst performing system when generating all six recommendations. However, our GQR (GPT-3) system always succeeds in generating six recommendations on all tested datasets. So, concerning RQ3, we can conclude that our GQR (GPT-3) system always produces recommendations for all ranks.

\begin{table}[ht!]
\centering
\begin{tabular}{l c c c c}
\toprule
\textbf{Model} & \textbf{One reccom.} & \textbf{Six reccom.} & \textbf{Average}\\
\midrule
\multicolumn{4}{c}{\textbf{Robust04}}\\
\midrule
\textit{System~1} & 92\% & 87\% & 5.39 \\
\textit{System~2} & 98\% & 86\% & 5.54 \\
Bhatia & 95\% & 89\% & 5.51 \\
GQR (Bloom)  & 99\% & 55\% & 4.69 \\
GQR (GPT-3) & \textbf{100\%} & \textbf{100\%} & \textbf{6} \\
RA-GQR (GPT-3) & \textbf{100\%} & \textbf{100\%} & \textbf{6} \\
\midrule
\multicolumn{4}{c}{\textbf{ClueWeb09B}}\\
\midrule
\textit{System~1} & 98\% & 92\% & 5.75 \\
\textit{System~2} & 99\% & 95\% & 5.85 \\
GQR (Bloom)  &  99\% & 63\%  & 5.1\\
GQR (GPT-3) &  \textbf{100\%} & \textbf{100\%} & \textbf{6} \\
RA-GQR (GPT-3) & \textbf{100\%} & \textbf{100\%} & \textbf{6} \\
\midrule
\multicolumn{4}{c}{\textbf{AOL long tail queries}}\\
\midrule
\textit{System~1} & 83\% & 71\% & 4.61 \\
\textit{System~2} & 91\% & 86\% & 5.37 \\
GQR (Bloom)  & 99\%  & 66\% & 5.15 \\
GQR (GPT-3) & \textbf{100\%} & \textbf{100\%} & \textbf{6} \\
RA-GQR (GPT-3) & \textbf{100\%} & \textbf{100\%} & \textbf{6} \\
\bottomrule
\end{tabular}
\caption{Percentage of queries for which at least one and all six recommendations (reccom.) are generated, and the average number of generated recommendations per query for different systems and datasets.}
\label{tab:long-tail}
\end{table}
\textbf{RQ4: Are query logs still bringing value in generative query recommendation?}

In our analysis, we show how a system comprised of a single component (the LLM) without using query logs can outperform more complex state-of-the-art query recommender systems that utilize query logs, such as System 1 and System 2. This suggests that query logs are not essential for building effective query recommender systems. Nevertheless, we recognize query logs as an important source of knowledge, as demonstrated by the performance of RA-GQR.

According to the \textit{Substitution} protocol, RA-GQR demonstrates significant improvements over all other systems tested, including GQR (GPT-3). As shown in Table 1 (\ref{tab:scs-subs}), RA-GQR enhances the SCS of GQR by approximately ${\sim}59\%$ on Robust04 and ${\sim}76\%$ on ClueWeb09B, compared to GQR (GPT-3). Additionally, Table 2 (\ref{tab:scs-subs}) shows that RA-GQR increases the NDCG@10 of GQR by roughly ${\sim}13\%$ on Robust04 and ${\sim}22\%$ on ClueWeb09B.

In the \textit{Concat} protocol, we observe similar improvements. Table 3 (\ref{tab:scs-conc}) confirms that RA-GQR nearly doubles the performance of GQR (GPT-3) on both datasets. Moreover, Table 4 (\ref{tab:ndcg-conc}) indicates that RA-GQR improves the NDCG@10 by about ${\sim}6\%$ on Robust04 and ${\sim}2\%$ on ClueWeb09B, compared to GQR (GPT-3). Overall, RA-GQR enhances the NDCG@10 by approximately ${\sim}11\%$ on Robust04 and ${\sim}6\%$ on ClueWeb09B with respect to the second-best system.

Therefore, in response to RQ4, while query logs are no longer essential, behavioral data still prove beneficial in enhancing the performance of generative query recommender systems.

\section{Conclusions and Future Work}\label{sec:conclusions}

In this work, we reframed the query recommendation task as a generative task presenting Generative Query Recommendation (GQR): a system that generates recommendations 
GQR generates query recommendations in a single forward pass without relying on long cascading pipelines and multiple systems to generate and re-rank the recommendations but leveraging a single component of an LLM and its knowledge.
GQR discontinues the use of users' data, such as query logs, resulting in an approach that is robust to long-tail queries, and cold-start recommendations.
Additionally, GQR does not require any task-specific architecture to be trained or fine-tuned because no data are involved in our query recommendation pipeline.

From a practical standpoint, this work has huge implications, as it proves that a query recommendation system that is competitive or even better than industry standards can be created with a pre-trained LLM that does not require any specific fine-tuning, but only a handful of curated prompts that can be obtained straightforwardly. This approach strongly democratizes the creation of query recommendation systems, and drastically reduces the time-to-market for new systems or existing systems that want to expand in new domains or markets.
We compared GQR against the previous state-of-the-art model presented by Bhatia et al. 2011 and two of the most popular search engines.

Analyzing the results, our approach outclasses the other systems on Simplified Clarity Score and Retrieval Effectiveness metrics and in our blind user study.
In terms of effectiveness, GQR (GPT-3) achieves an improvement of up to ${\sim}4\%$ of NDCG@10 score in Robust04 and ClueWeb09B w.r.t. the best competitor system.
Regarding the Simplified Clarity Score, GQR (GPT-3) outperforms the other systems on average, and it turns out to be the system with the lowest standard deviation, demonstrating that it has the most stable performance along the recommendations generated in the various ranks.
We then propose an enhanced GQR proposing Retrieval Augmented GQR, which is capable of retrieving similar facts from a query log to compose its prompt dynamically, which further improves its performance.
Indeed, it obtains improvements of ${\sim}11\%$ and ${\sim}6\%$ in Robust04 and ClueWeb09B with respect to the second-best competitor.
In our blind user study, we recorded 60\% of preferences for our system, indicating that our method is not only superior in terms of metrics but also capable of generating engaging queries for users.

From our methodological perspective, we have proven that general-purpose LLMs can compete or even outperform other established methods for a fundamental search task.

In future work, we plan to delve deeper into the customization of prompts for query recommendations, investigating how nuanced modifications can enhance the effectiveness of our system. Additionally, we aim to extend our exploration to include exploratory queries, assessing how different configurations of prompts can improve engagement and relevance in more complex or open-ended information-seeking tasks.

\begin{acks}
This work is supported by the Spoke ``FutureHPC \& BigData" of the ICSC – Centro Nazionale di Ricerca in High-Performance Computing, Big Data and Quantum Computing, the Spoke “Humancentered AI” of the M4C2 - Investimento 1.3, Partenariato Esteso PE00000013 - "FAIR - Future Artificial Intelligence Research", SERICS (PE00000014), IR0000013 - SoBigData.it, funded by European Union – NextGenerationEU, the FoReLab project (Departments of Excellence), and the NEREO PRIN project funded by the Italian Ministry of Education and Research Grant no. 2022AEFHAZ. This work was carried out while Andrea Bacciu was enrolled in the Italian National Doctorate on Artificial Intelligence run by the Sapienza University of Rome.
\end{acks}
\balance

\newpage
\bibliographystyle{ACM-Reference-Format}
\bibliography{biblio}

\end{document}